\documentclass[fleqn,twoside]{article}
\usepackage{espcrc2}

\usepackage{graphicx}

\newcommand{\AmS}{{\protect\the\textfont2
  A\kern-.1667em\lower.5ex\hbox{M}\kern-.125emS}}

\def\simle{
    \mathrel{\rlap{\raise 0.511ex
        \hbox{$<$}}{\lower 0.511ex \hbox{$\sim$}}}}
  
\title{Bayesian approach to the first excited nucleon state in 
lattice QCD
 \thanks{This work is supported by the Supercomputer Project No.85 
	       (FY2002) of High Energy Accelerator Research 
	       Organization (KEK).
	       S.S. thanks for the support by JSPS Grand-in-Aid for 
	       Encouragement of Young Scientists (No.13740146).	       
	       } 
}

\author{S. Sasaki\address[DPUT]{Department of Physics, University 
                                     of Tokyo, Tokyo 113-0033, 
Japan}, %
          K. Sasaki\addressmark[DPUT], 
	  T. Hatsuda\addressmark[DPUT], 
	  M. Asakawa\address[DPKU]{Department of Physics, 
	                           Kyoto University,
	                           Kyoto 606-8502, Japan}}       
\begin{document}

\begin{abstract}
We present preliminary results from the first attempt to
reconstruct the spectral function in the nucleon and $\Delta$ channels
from lattice QCD data using the maximum entropy method (MEM). 
An advantage of the MEM analysis is to enable us to access 
information of 
the excited state spectrum.  Performing simulations on two lattice 
volumes, we confirm the large finite size effect on 
the first excited nucleon state in the lighter quark mass region.
\vspace{-15pt}
\end{abstract}

\maketitle

In quantum field theory, the spectral functions (SPFs) of 
the two-point correlation function is expected to expose
rich physical information which is not only for the ground state,
but also for excited states. 
Nevertheless, the reconstruction of SPF, $A(\omega)$ from given Monte 
Carlo data of 
the Euclidian time correlator: $G(t)=\int d\omega A(\omega) 
\exp (-\omega t)$ is a typical ill-posed problem.
The maximum entropy method (MEM) is a useful method 
to circumvent such ill-posed problem by making a statistical
inference of the most probable image of $A(\omega)$ based on Bayesian
statistics.

Recently the MEM analysis is widely employed on various
problems in lattice simulations after the first success
in our research area \cite{{Naka99},{Asak01}}. 
As for the light hadron spectroscopy, 
the CP-PACS collaboration analyzed their own high-precision
quenched lattice QCD data using the MEM. They show the
reliability of the MEM through checking consistency between
the standard analysis and the MEM analysis after the continuum 
extrapolation \cite{Yama02}.
However above applications were carried out only for mesonic hadrons.
In this article, we apply the MEM analysis to lattice QCD data for 
both spin-1/2 and spin-3/2 baryons in order to study the
excited state spectrum of baryons.

We are interested in a long standing puzzle regarding 
the excited state spectrum of the nucleon, 
namely, the level order of the positive-parity excited nucleon
$N'(1440)$ and the negative-parity nucleon $N^*(1535)$. 
The pattern of the level order between positive and negative-parity
excited states can be found universally in the $\Delta$,
$\Sigma$ and flavor-octet $\Lambda$ channels. 
Recent quenched lattice QCD studies show that the wrong ordering 
between $N'$ and $N^*$ actually happens in the relatively {\it 
heavy-quark}
mass region~\cite{Sasa02}.  
Thus, we address a serious question whether 
or not the level switching between $N'$ and $N^*$ would occur in the
{\it light-quark} mass region. Finding of this possibility might be 
directly
connected to the understanding of the mysterious Roper resonance.

We remark that the simulation for the light-quark mass requires
large lattice size since the ``wave function'' of 
the bound state enlarges as the quark mass decreases.
Once the ``wave function'' is squeezed due to the small volume,
the kinetic energy of internal quarks raises and thus
the total energy of the bound state should be pushed up.
This is an intuitive picture for the finite size effect on the 
mass spectrum. Such effect is expected to become serious for 
the (radial) excited state rather than the ground state.
Indeed, ref.\cite{Sasa02} reported that
the $N'$ mass in the 
light-quark region 
is significantly heavier than the mass 
extrapolated from the
heavy quark region.
Their lattice simulation was performed on relatively 
small volume ($L\approx 1.5{\rm fm}$).

%
%
\begin{figure}[t]
\includegraphics[width=68mm]{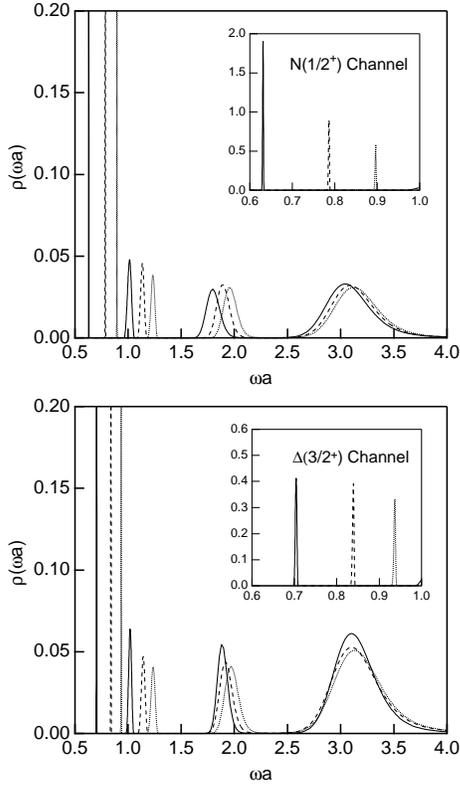}
\vspace{-30pt}
\caption{
Spectral functions in the $N(1/2^+)$
channel
and the $\Delta(3/2^+)$ channel. }
\label{fig:SpfNucl}
\vspace{-20pt}
\end{figure}

To study the finite size effect, numerical simulations are performed 
on two lattice sizes $16^3 \times 32$ and $24^3 \times 32$. 
We generate quenched QCD configurations with the standard 
single-plaquette 
action at $\beta = 6.0$ ($a^{-1}\approx 1.9 {\rm GeV}$).
The quark propagators are computed using the Wilson 
fermion action at four values of the hopping parameter $\kappa$,
which cover the range $M_{\pi}/M_{\rho}\approx 0.69-0.92$. 
Our preliminary results are analyzed on 352 configurations 
for the smaller lattice ($L\approx1.5{\rm fm}$) 
and 300 configurations for the larger lattice ($L\approx2.2{\rm fm}$).

We use the conventional interpolating operators, 
$\varepsilon_{a b c}(u^{T}_{a}C\gamma_{5}d_{b})u_{c}$
for the nucleon and $\varepsilon_{a b 
c}(u^{T}_{a}C\gamma_{\mu}u_{b})u_{c}$ 
for the $\Delta$ respectively. Correlators constructed from those 
operators are supposed to receive contributions from both positive and
negative-parity states. More details of the parity projection are 
described in ref.\cite{Sasa02}. 

For the $\Delta$ 
correlator $G^{\Delta}_{\mu \nu}$, we need the spin projection to 
extract desired spin state since there are contributions from 
both $J=3/2$ and $1/2$ states. At zero spatial momentum, 
the $\Delta$ correlator with spatial Lorentz indices is 
expressed in the form
%
%
\begin{equation}
G^{\Delta}_{i j}(t) = \left(\delta_{i j}
-\frac{1}{3} \gamma_{i}\gamma_{j}\right) G^{\Delta}_{3/2}(t)
+ \frac{1}{3} \gamma_{i}\gamma_{j} G^{\Delta}_{1/2}(t),
\end{equation}
where contributions from negative-parity states are omitted for 
simplicity.
Defining ${\bar G}_{i j}= \frac{1}{4}{\rm Tr}
(\gamma_{i}\gamma_{j}G^{\Delta}_{j i})$, one can obtain each 
amplitude; 
$G^{\Delta}_{3/2}$ and $G^{\Delta}_{1/2}$ from appropriate 
combinations of
${\bar G}_{i j}$. In this article, we only analyze the $J^{P}=1/2^{+}$
part of the $N$ correlator and the $J^{P}=3/2^{+}$ part of
the $\Delta$ correlator.

We define the dimensionless SPF for baryons through 
$A(\omega)=\rho(\omega)\omega^5$ \cite{Asak01}. 
In the continuum perturbative QCD, 
the asymptotic (renormalized) 
value of $\rho^{\rm ren}(\mu)$ for the large renormalization scale 
$\mu (\gg1{\rm GeV})$ 
can be evaluated within the one-loop level 
%
%
\begin{equation}
    \rho^{\rm ren}(\mu)=
\frac {c_{1}}{(2\pi)^4} \left(1+c_{2}
    \frac{\alpha_{s}(\mu)}{\pi}\right).
\end{equation}
Here the value of $c_{1}=5/128\;(1/10)$ and $c_{2}=71/12\;(52/45)$ 
for our used 
$N$ ($\Delta$) operator in the $\overline{\rm MS}$-scheme can be 
retrieved 
from the number quoted in ref.\cite{Jami88}.
In our MEM analysis, the prior knowledge (default model; $m_{0}$) 
is deduced from $\rho^{\rm ren}(\mu)/(Z_{3q}^{\rm latt}(\mu, a))^2$ 
at $\mu = a^{-1}$. $Z_{3q}^{\rm latt}$ 
denotes the renormalization factor of the specified
baryon interpolating operator on the lattice.
For the $N$ operator, the tadpole improved perturbative result
is give by $(1-0.73 \alpha_{V})/8$ in the chiral limit \cite{Lepa93}. 
We roughly choose $m_{0}= 4\times 10^{-3}$ for $N$ and 
$8 \times 10^{-3}$ for $\Delta$ according to the above procedure.
The results are insensitive to the variation of $m_{0}$ around this 
choice.

The MEM analysis is performed by using data up to the
half of the temporal extent except for the source location.
Most of the results 
presented here use $N_{\omega}\sim O(600)$ and $\omega_{\rm 
max}a\approx 2\pi$. The detail procedure of the MEM 
can be found in \cite{Asak01}.

%
%
\begin{figure}[t]
\includegraphics[width=70mm]{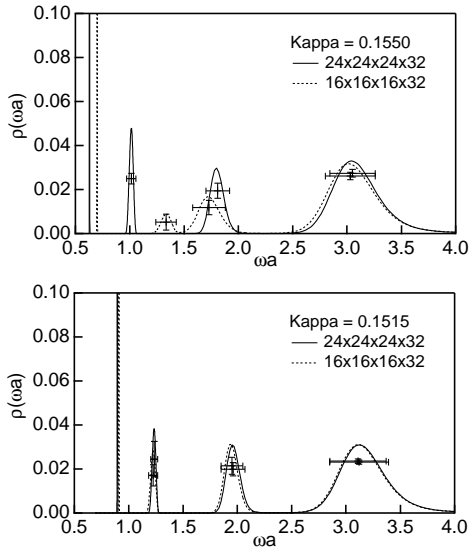}
\vspace{-30pt}
\caption{
Study of the finite size effect in the nucleon channel
at $\kappa=0.155$ and 0.1515.
}
\vspace{-30pt}
\label{fig:FVSpf}
\end{figure}
%

%
%
\begin{figure}[t]
\includegraphics[width=68mm]{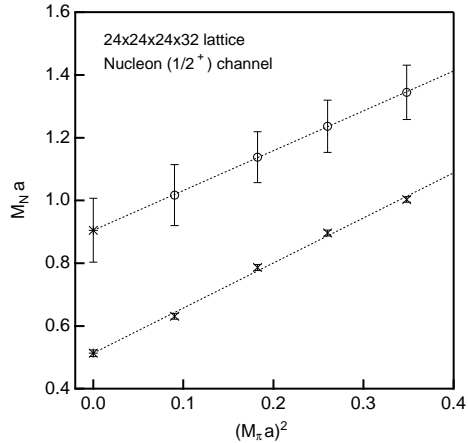}
\vspace{-30pt}
\caption{The ground state ($\times$) and the first excited state 
($\circ$) masses versus the pion mass squared in the nucleon channel.}
\vspace{-20pt}
\label{fig:ChMass}
\end{figure}

In Fig.1, we show SPFs in the nucleon and
$\Delta$ channels on the larger volume ($L\approx2.2{\rm fm})$. 
The solid, dashed and dotted lines are for $\kappa$=0.155, 0.153 
and 0.1515 respectively.
In both channels, there are one sharp peak, one broad peak and two 
bumps at each $\kappa$. 
Two bumps at each $\kappa$ have large overlap with the ones
at different $\kappa$ while the peak positions are relatively 
shifted to the left as $\kappa$ increases toward the chiral limit.
Those states might be the unphysical bound states 
of a physical quark and two doublers, which have been found 
in the mesonic case \cite{Yama02}.
The additive simulations at different 
lattice spacing are necessary to confirm this speculation.

In order to examine the finite size effect, we compare
SPFs of the nucleon on the larger volume 
with the ones on the smaller volume at $\kappa=0.155$ and 0.1515
in Fig.2. The crosses 
on each peak or bump represent the statistical significance of SPF 
obtained  by the MEM.
We see a large finite volume effect on the second peak 
for the lighter quark ($\kappa=0.155$) as compared with the 
first peak and two bumps.
It indicates that the (physical) excited state is significantly
affected by the finite size effect in comparison to the ground state.

We plot the $N$ and $N'$ masses, which correspond to 
the peak positions of first two peaks, as a function of the pion mass 
squared 
in Fig.3. The errors are estimated by the jackknife method.
We mention that the $N$ masses are quite consistent with the ones 
determined from the single exponential fits. 
In addition, for the $N'$, the heaviest two points are
also good agreement with the corresponding results of 
ref.\cite{Sasa02}.
Taking a simple linear extrapolation in Fig.3, 
we find $M_{N}$=0.51(1) and $M_{N'}$=0.90(10) in lattice units.
Our results may be compared with the previously 
published results 
for the $N^*$ 
at the same lattice spacing;
$M_{N^*}=0.85(5)$ \cite{Sasa02}
for lattice size $L\approx 1.5 {\rm fm}$ with domain wall fermions
and 0.89(2) \cite{Rich02} for lattice size $L\approx 2.2 {\rm fm}$ 
with 
clover fermions.
We find that the level spacing between $N^*$ and 
$N'$ reduces significantly in the chiral limit. 
However the level switching between them might not happen 
in lattice simulations
with $L \simle 2.2 {\rm fm}$.

We have applied the maximum entropy method to 
lattice QCD data for both spin-1/2 and spin-3/2 baryons to study 
the positive-parity excited state spectrum. We succeeded in 
extracting SPFs for baryons as well as mesons. 
Based on the systematic analysis utilizing two lattice sizes, 
we confirmed the large finite size effect on the first excited 
nucleon state in the light quark mass region originally pointed out in 
ref.\cite{Sasa02}. 


\end{document}